\documentclass[USenglish,oneside,twocolumn]{article}

\usepackage[utf8]{inputenc}
\usepackage[big]{dgruyter_NEW}

\usepackage{graphicx}
\usepackage{times}
\usepackage{balance}
\usepackage{float}
\usepackage{url}

\usepackage[caption=false]{subfig}
\usepackage{color,soul}
\usepackage{makecell}
\usepackage{amsfonts}
\usepackage{amssymb}
\usepackage{tabulary}

\usepackage{booktabs} 
\usepackage{makecell}

\DOI{foobar}

\cclogo{\includegraphics{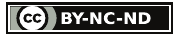}}
  
\begin{document}
 
\author*[1]{Kopo M. Ramokapane}

\author[2]{Anthony C. Mazeli}

\author[3]{Awais Rashid}

\affil[1]{Bristol Cyber Security Group, University of Bristol, marvin.ramokapane@bristol.ac.uk}

\affil[2]{Security Lancaster Institute, Lancaster University, tonymazeli@yahoo.com} 

\affil[3]{Bristol Cyber Security Group, University of Bristol, awais.rashid@bristol.ac.uk}

\title{\huge Skip, Skip, Skip, Accept!!!: A Study on the Usability of Smartphone Manufacturer Provided Default Features and User Privacy}

\runningtitle{Article title}

  \begin{abstract}
{
	Smartphone manufacturer provided default features (e.g., default location services, iCloud, Google Assistant, ad tracking) enhance the usability and extend the functionality of these devices. Prior studies have highlighted smartphone vulnerabilities and how users' data can be harvested without their knowledge. However, little is known about manufacturer provided default features in this regard---their usability concerning configuring them during usage, and how users perceive them with regards to privacy. To bridge this gap, we conducted a task-based study with 27 Android and iOS smartphone users in order to learn about their perceptions, concerns and practices, and to understand the usability of these features with regards to privacy. We explored the following: users' awareness of these features, why and when do they change the settings of these features, the challenges they face while configuring these features, and finally the mitigation strategies they adopt. Our findings reveal that users of both platforms have limited awareness of these features and their privacy implications. Awareness of these features does not imply that a user can easily locate and adjust them when needed.
	Furthermore, users attribute their failure to configure default features to hidden controls and insufficient knowledge on how to configure them. To cope with difficulties of finding controls, users employ various coping strategies, some of which are platform specific but most often applicable to both platforms. However, some of these coping strategies leave users vulnerable.
}
\end{abstract}

  \journalname{Proceedings on Privacy Enhancing Technologies}
\DOI{10.2478/popets-2019-0027}
  \startpage{1}
  \received{2018-08-31}
  \revised{2018-12-15}
  \accepted{2018-12-16}

  \journalyear{2019}
  \journalvolume{2019}
  \journalissue{2}

\maketitle

\section{Introduction}
\label{Label:Introduction}
\vspace{-3mm}
Mobile phone manufacturers are constantly developing new and innovative features to leverage advancements in smartphone technology while maintaining relevance in the competitive global market~\cite{raento2009smartphones,dufau2011smart,yarkoni2012psychoinformatics}. By default, for instance, both Apple iOS and Google Android platforms are rolled out with an enabled~\emph{ad identifier} to help users with personalized adverts. There is also~\emph{location} which is primarily used by other apps (e.g., Maps) to provide the geographical location of the device.

Controversially, platform providers usually use \textit{opt-out} notices for personal data collection though it has been established that users are known to keep defaults~\cite{hadar2018privacy}. An \textit{opt-out} notice is an agreement that requires a user to make an informed decision and take actions to deny consent. By default, a user has given consent. Research knows little about smartphone default features or how they are perceived by users with regards to privacy\textemdash whether users are even aware of these features or the privacy implications of leaving them unchanged. Like other privacy-related control settings (e.g., App permissions), lack of awareness of manufacturer-provided default features (MPDFs) poses various privacy risks, e.g., accidental data disclosures resulting in discomfort and regrets for users~\cite{lin2012expectation, felt2012android, almuhimedi2015your}.

Research has highlighted users' understanding and behavior pertaining to app permissions on both iOS and Android platforms~\cite{iachello2005developing,chin2012measuring,felt2012android,felt2012ve,liu2014reconciling, Tsai205132}. However, these have mainly focused on the permissions of downloaded or third-party applications. No work has considered permissions of default features. MPDFs are different from App permissions in that they are `features' rolled out with the smartphone platform, and users are likely to encounter them -- from a privacy settings perspective -- only once in their usage of smartphone, that is, during initial setup. However, prior studies have shown that users pay no attention to permissions or settings~\cite{felt2012android}. Unlike application developers, platform providers are usually in a much more powerful position and, before installing third-party applications, users already show trust in these platforms~\cite{Lena2014}. During initial smartphone setup, users are likely to \textit{skip} and \textit{accept} the options on offer to start utilizing their newly acquired devices. However, this behavior may lead to users sharing their personal data with the provider unknowingly. For instance, when setting up a new phone, iPhone users are usually encouraged to provide their Apple ID credentials to continue with the setup and enable iCloud. However, the implications of this action are not made clear, i.e., this will automatically sync their mobile phone data with iCloud, something a user may not wish to do. Other features rolled out in a similar fashion include \textit{location-based} services, \textit{ad tracking} and sharing of \textit{analytical data} with the platform provider. Users are then expected to know that these features are enabled. Moreover, if they wish to disable them, it is assumed they will know where they are located and be able to configure them. 

To date, no work has focused on understanding how users perceive MPDFs and the usability of configuring them concerning privacy. Without such understanding, researchers and platform designers can neither prioritize areas on which to focus nor develop usable solutions that empower users to manage their privacy effectively.

To the best of our knowledge, we are the first to conduct a study on MPDFs\textemdash users' perceptions towards them, their usability and implications on users' privacy. We report on a qualitative interview and task founded analysis, based on 27 iOS and Android users. Using thematic analysis and task-based exercises, we answer the following research questions: (1) how aware are smartphone users of MPDFs with regards to privacy? (2) In which context do smartphone users adjust their default settings? (3) What specific problems (if any) do smartphone users encounter? And finally, (4) what coping strategies (if any) do they adopt to overcome usability problems when configuring their privacy settings for MPDFs?

The novel contributions of our work are as follows:
\vspace{2pt}

\textbf{We provide a usability study of iOS and Android MPDFs, highlighting users' awareness levels and attitudes.}\\
We find that most users are not aware of the privacy implications of MPDFs, they mostly learn about the risk from other people. Some users treat MPDFs as part of the platform and find no reason to adjust them. Being aware of MPDFs does not necessarily imply that users can find and configure them when they want. When setting up new devices, most users prefer to leave MPDF settings in the default mode in order to enjoy the features of their new phones quickly, mostly not returning to change them. Moreover, while users are aware of the privacy implications of MPDFs, some are unconcerned and may choose not to change them.

\textbf{We provide evidence on what motivates smartphone users to configure their MPDFs during initial setup or usage. We also uncover their challenges when adjusting MPDFs and the mitigation strategies they employ when faced with such challenges.}
Our results suggest that the decision to adjust MPDFs is not only affected by users' privacy concerns, but it is also dependent on their proficiency level. When they know where to look, they are more likely to adjust their settings. App prompt requests do not only raise awareness of MPDFs but also motivate users to change them. Failing to find where MPDF settings are located is a common challenge for many users. These failures lead to a range of coping strategies, including users resigning themselves to having their data being utilized by platform providers, relying on others to help them or searching for information online.

\textbf{We elicit users' mental models pertaining to MPDFs and privacy as well as guidelines for future designs for MPDFs.}
We identify 11 mental models that play a role in users' decision to alter MPDF settings, how they carry out the configuration task and their coping strategies in response to failures. These coping strategies vary: some depend on the platform while others are influenced by users' motivations. Our study also highlights that users want MPDFs to be grouped under a single menu so that they are easy to locate when they need to be adjusted. Furthermore, to raise awareness, platforms should notify users when such features collect or share data, and be more transparent about what data is being collected and how it is being utilized.
\vspace{-7mm}

\section{Background}
\label{Label:Background}
\vspace{-3mm}
All smartphone features have initial settings or configurations that a user is required to modify per need or desire. To change these settings (during the initial setup of the device) – while one is still excited about getting a new device \textemdash a user is required to understand the consequences of leaving them unchanged. However, because of the time and effort required, most users often \emph{skip} or \emph{continue} to accept the defaults without any understanding or awareness of the future implications of leaving them unchanged~\cite{gross2005information,acquisti2015privacy}. Default settings can lead to undesired privacy implications and affect users' lives~\cite{watson2015mapping}.

When Android users set up their phone (Android 6.0), they are first asked to select their preferred language and connect to the network. After being given a chance to transfer data from another phone, they are required to either create a new Google account or login with an existing one. If a user decides to sign in with their Google account or create a new one, then by default, they are opted into backing up their data, sharing location, improving location services, and sending usage and diagnostic information to Google. The last step before the home screen requires them to opt out of using Google Assistant. These steps are different when a user decides not to use their Google account or connect to the network. Moreover, some Android devices may also include the device manufacturer settings as part of this process.

To set up an iPhone (iPhone 6), a user is first required to select their preferred language and region. The next step is to connect to the network and insert a SIM card if they have not already done so. A user is then given a chance to enable location services, set up their Touch ID and passcode. The subsequent step requires a user to choose between setting up their phone as a new device, restoring from iTunes or iCloud backup, and moving data from an Android device. If the user decides to set up the phone as a new device, they will be asked to create or login with their Apple ID and then asked to accept terms and conditions about using the platform. Users are then asked to set up their apple pay, iCloud keychain, and Siri. The last stage requires users to decide whether they want to share diagnostic and App analytics data with Apple. These steps are different if the user decides against setting the device as a new one. 

While there are many MPDFs enabled by default, users are usually shown location services, backup features, and diagnostic or analytics data settings. Table~\ref{tab:defaultfeatures} shows a list of some of these features. Most MPDF settings are hidden behind many screens; users are usually required to navigate through several screens to find these settings (e.g., configuring \emph{Ads identifier}). 

Some MPDFs cannot be disabled by a single control, for instance, location. Android users have to disable \emph{location} under settings but this does not sufficiently restrict location tracking. To fully restrict tracking, they are further required to disable this under \emph{activity tracking}. In iOS \emph{location services} can also be tracked through \emph{significant locations} feature.

For Android users, these features can be more confusing because they are sometimes obliged to also decide on features which have been introduced by the device manufacturer, for instance, using \emph{smartswitch} on Samsung or having to configure a second location feature on Motorola X. This is not only cumbersome but also confusing for users because there is never a clear distinction between the two – platform and device features. Our study focuses on platform features. 

MPDFs can be very useful to users, for instance, using Apple ID to set up one’s iPhone may save them from permanent data loss or enabling location (Android Manager) may help a user delete their data remotely without the device. Moreover, based on the information used during configurations or setup, users sometimes infer default settings as implied recommendations from the provider~\cite{acquisti2015privacy}. For most users, leaving settings unchanged may be perceived as helping the provider to improve their device or platform because of the language used during smartphone setup. In most cases, this gives users an illusion that they understand the consequences of their decisions. However, recent events~\cite{Facebook2017teen, Facebook2018Analytica,venkatadri2018investigating} have revealed how service providers can misuse data, for instance, information that is being collected can be aggregated and be used to infer to someone’s behavior patterns. Currently, both Google and Apple do not have a comprehensive list of their MPDFs or where users can go in order to change these settings.
\vspace{-3mm}

\begin{table}[!ht]
	\small
	\centering
	\caption{Android and iOS commonly enabled MPDFs. $\checkmark$ marks the features we investigated.}
	\renewcommand{\arraystretch}{0.8}
	\begin{tabular}{l l}
		Android MPDF & Apple iOS MPDF\\
		\hline
		\hline
		- Location $\checkmark$ & - Location services $\checkmark$ \\
		- Improve location & - Siri and search\\
		- Diagnostic \& usage data $\checkmark$ & - Analytics $\checkmark$ \\
		- Ads tracking $\checkmark$  & - Website Tracking in safari\\
		- Google Now ``Assistant'' & - Auto fill\\
		- Activity tracking & - Safari Camera \& Mic Access\\
		- Google Drive backup & - Show Parked Location\\
		- Photo Geo tag &	- Significant/frequent locations\\
		- Voice Tracking & - Raise to Wake\\
		- Google keyboard & - iCloud\\
		(Usage statistics \& snippets) & - Ads tracking or identifier $\checkmark$ \\
		\hline
	\end{tabular}
	\label{tab:defaultfeatures}%
\end{table}
\vspace{-10mm}

\section{Related Work}
\label{Label:RelatedWork}
\vspace{-3mm}
There is a wealth of prior work on the security and privacy of smartphones. Researchers have explored users' awareness and concerns concerning privacy, studied privacy leaks on both Android and iOS applications, the usability challenges of privacy controls, and proposed solutions to help users manage privacy. In this section, we discuss how existing works have influenced the design of our study and the gap with regards to the understanding of MPDFs and user privacy.
\vspace{-4mm}

\subsection{User Awareness and Privacy Concerns}
\vspace{-3mm}
Several studies have found that smartphone users are more concerned about their privacy only after realizing that their decisions have put them at risk of data exfiltration through third-party apps~\cite{jung2012short,felt2012android,thompson2013s}. Shih et al. showed that the likelihood of users disclosing their private information was higher when they were confused and uncertain, or the purpose for such disclosure was unclear~\cite{shih2015privacy}. Others have found that users are often unaware of the number of permissions they grant through privacy controls~\cite{almuhimedi2015your,balebako2013little,felt2012android}. Permission model studies~\cite{felt2012ve,egelman2013choice,shklovski2014leakiness} also reveal that users are mostly unaware of how apps access protected resources and how that can be regulated. Moreover, Micinski et al. have shown that the majority of access requests for features such as location information sometimes happen without the user's knowledge~\cite{Micinski2017}. Felt et al. and Kelley et al. cited complex user interfaces and lack of understanding of app permissions by users as the main contributors to misconfiguring app permissions~\cite{felt2011android,kelley2012conundrum}. They also highlighted that few users actually peruse privacy terms and conditions. Despite all these efforts, users' awareness and privacy concerns of MPDFs have not been explored previously. We tailored our study to understand users' awareness of privacy-related default features.

King interviewed 24 iPhone and Android users about their privacy preferences and expectations, and found that users were far less concerned about sharing location compared to other types of information available through their platforms~\cite{king2012come}. She also found that most App developers defy users' privacy expectations. However, King's work is restricted to third-party apps and does not consider MPDFs.
\vspace{-4mm}

\subsection{Android and iOS Privacy Leakages}
\vspace{-3mm}
Prior research on iOS and Android privacy leakages focuses mostly on detecting privacy leaks through third-party applications~\cite{zhou2011taming,hornyack2011these,kim2012scandal,Clint2012,enck2014taintdroid}. Egele et al.~\cite{egele2011pios} and Agarwal et al.~\cite{agarwal2013protectmyprivacy} have also analyzed third-party Apps for iOS. Egele et al. found that more than half of applications were found to leak the unique ID of the device which provided detailed information about the users' mobile activities. While these studies focus on preventing such leaks, they only consider downloaded apps and not MPDFs. 
\vspace{-5mm}

\subsection{Privacy Preserving Solutions}
\vspace{-3mm}
Others have focused on tools and applications that can be used to help users manage their privacy settings on smartphones. Works reported in~\cite{jedrzejczyk2010impact,liu2016follow,liu2016privacy} have
explored ways to nudge or alert users to configure their privacy settings in order to make them aware when and how applications access their data. In order to give users more control, several studies, e.g.,~\cite{hornyack2011these,shebaro2014identidroid} went on and designed tools that dynamically block runtime permission requests, and those that enable users to deny data to applications or to substitute users' data with fake data. Other studies have proposed crowdsourcing approaches to help users decide which permissions to disable~\cite{ismail2015crowdsourced,ismail2017permit}. Liu et al. suggested the employment of profile-based personalized privacy assistant (PPA) to alleviate users' configuration burdens on both iOS and Android~\cite{liu2016follow,liu2016privacy}. Tsai et al. confirmed that users are better suited to performing permission management tasks using tools such as TurtleGuard than with the default permission manager~\cite{tsai2017turtle}. However, these studies are limited to Android smartphone users only and focus on app permissions rather than MPDFs. Our study examines some of the default features of both Android and iOS platforms. In our study, we sought to investigate from users whether similar tools for MPDFs exist and if users employ them as coping strategies.
\vspace{-4mm}

\section{Methodology}
\label{Label:Methodology}
\vspace{-3mm}
To answer our research questions, we invited 27 Android and iOS smartphones users to take part in an interview and task-based study between June and August 2017.
\vspace{-4mm}

\subsection {Ethical Consideration} 
\vspace{-3mm}
Our study was approved by the relevant Institutional Review Board (IRB) process before any research activity began. We obtained informed written consent from all participants to take part in the study and to have the interviews audio recorded.
\vspace{-4mm}

\subsection {Participant Sampling}
\vspace{-3mm}
Most of our participants were recruited through advertised posters around our institution while the rest were recruited through our existing professional networks and word of mouth. To reduce biases, we advertised our study as a usability study for mobile platforms. Interested respondents were invited to complete an online questionnaire tailored towards screening suitable participants who could be invited to take part in the study. We asked them to select which, if any, mobile phone platform they used, how long they had been using it, whether they had ever set up a new mobile phone before and how recent was it. We also asked them to rate how confident they were about configuring or changing their mobile phone settings. There were demographic information questions as well (i.e., age, gender, education, and profession).

We set out to identify a group of between 25 to 30 participants who used either iOS or Android phones. We were looking for respondents meeting the following criteria: (1) stated that they had set up or configured a new smartphone before, (2) had moderate to sufficiently high smartphone usage, (3) possessed reasonable (self-rated) capabilities in configuring smartphones, and (4) had been using them for a period of at least 6 months to 2 years. We also asked them to state whether privacy played any significant role in their choice of smartphone. Apart from this, the screener was also used to balance the demographics such as age, gender and the type of smartphone they were using.

During the course of 4 weeks, we received 52 respondents and invited 30 of them to take part in the study. Out of the 30 participants who were invited, only 27 showed up and completed the study. Three participants either failed to turn up or declined the invitation. Out of the 27 who took part in the study, 13 (5 males and 8 females) were Android users while 14 (6 males and 8 females) were iOS users. Table~\ref{tab:Demographics} shows a summary of the demographics of our participants.

\begin{table}[!ht]
	\small
	\centering
	\caption{ Summary: Study Demographics.}
	\renewcommand{\arraystretch}{0.9}
	\begin{tabular}{l c}
		\hline
		\textbf{} &  \textbf{No. of participants}  \\
		\hline
		\textbf{Gender} & \textbf{}  \\
		Male & 11 \\
		Female & 15 \\
		Preferred not to say & 1 \\
		\hline
		\textbf{Age} & \textbf{}  \\
		18 - 20 &  7  \\
		21 - 25 &  10  \\
		26 - 30 &  5  \\
		31 - 40 &  3  \\
		45 + &  2 \\
		\hline
		\textbf{Education} & \\
		High school/College course & 7\\
		Bachelors & 11 \\
		Masters & 7 \\
		PhD & 2 \\
		\hline
		\textbf{Employment status} & \\
		Unemployed/Retired & 1 \\
		Full time & 8 \\
		Part-time & 3 \\
		Student & 14 \\
		Preferred not to say & 1 \\
		\hline
		\textbf{Smartphone category} & \\
		Android & 13\\
		iOS & 14 \\
		\hline
		\textbf{Smartphone users per gender} & \\
		Android & (Males: 5, Females: 8)\\
		iOS & (Males: 6, Females: 8)\\
		\hline
	\end{tabular}%
	\label{tab:Demographics}%
\end{table}
\vspace{-7mm}

\subsection{Study Procedure}
\vspace{-3mm}
Participants who were selected for the study were invited for full interviews. Each session began by taking the subject's consent to take part in the study and allow audio recording. For the main part of the interview, we employed two methods: informal cognitive walkthrough and think aloud-approach. Cognitive walk-through is a usability inspection method modeled after the software engineering practice of code walkthroughs~\cite{nielsen1994usability, Blackmon2002, Norgaard2006}. Each step of the user's problem-solving process is observed to see if the user's goals and memory for actions can be assumed to lead to the next correct action~\cite{Nielsen1994}. We used the think-aloud approach to capture these thoughts and reasons behind their decisions during tasks. The think-aloud method we adopted was the coaching method where participants are probed, prompted, and encouraged to describe their actions while completing tasks~\cite{olmsted2010think}. We adopted these two methods to elicit and understand users' mental models with regards to configuring MPDFs. To understand users' practices and minimize missing out on other issues related to MPDFs, during the exercises we asked users to report on the challenges they generally face and the mitigation strategies they adopt to alleviate such challenges.

We chose two popular brands of smartphones: Motorola X ($2^{\text{nd}}$ Generation) running Android version 6.0 Marshmallow and iPhone 6 running iOS version 10.3.2. These were the latest versions at the time of the study (2017, August). At the beginning of each session, each participant was given a device running the OS that they stated that they were using and comfortable with, reset to default settings. The participant was asked to complete four (4) different tasks depending on their OS platform, the whole session lasting between 30 and 45 minutes in total. On average each participant took 7 minutes to complete each task together with answering related questions. Before completing each task, participants were asked some questions related to the feature they were about to configure. This was done to clarify if the participants were familiar with the feature being discussed. However, to avoid priming participants, we excluded questions which could affect their actions while completing the task. For instance, if the task was to disable or limit~\textit{ad tracking}, then we would ask participants how they felt about browsing activities being monitored for targeted advertising purposes rather than asking if they knew such features are sometimes enabled by default. Section~\ref{sec:studytasks} explains these tasks in detail.

For both sets of users (iOS and Android), we started by asking them to disable~\textit{Location service} for all apps, a default feature (e.g., location) which is required by other downloaded apps to run, and then~\textit{ad tracking}. The order of these tasks was intentional; we started with the location feature because it is a commonly recognizable and requested mobile service. After these three tasks, each participant was asked to complete a task which was specific to their platform; Android users were asked to restrict sharing~\textit{Usage and Diagnostic Reports} while iOS users were asked to disable~\textit{Analytics Data} sharing. As each participant completed each task, we observed and noted their actions and decisions. Where necessary, we asked them questions for clarity. At the end of each session, we asked each participant to share their views on the study and default features before we compensated them with a small payment in Amazon vouchers (\$10.00) for taking part in the study.

We considered observing users configuring out of box devices. However, during our pre-study pilot, some users noted discomfort using their personal accounts to set up our devices. When we gave them accounts to use, some users intentionally skipped some settings and stated that they deliberately did so because they were not setting up their own phone. This may have negatively influenced our result, and they would not reflect how users indeed set up their mobile phones. Checking participants' settings on their own mobile devices would not be in line with ethical research practices either. Hence, we relied on their report.
\vspace{-4mm}

\subsubsection{Privacy Task-based Exercises}
\label{sec:studytasks}
\vspace{-3mm}

In order to understand the usability of configuring privacy for MPDFs, we designed 3 tasks applicable to both iOS and Android and one task specific to each platform. 

These tasks were based on privacy control configurations consistent with the default setup procedures of either iOS or Android when using the smartphones provided to our participants. The intention was to replicate real task-based scenarios like what the participant would normally encounter when configuring their smartphones in the first instance or during usage. Our focus was on privacy-related default settings which were present in both platforms or related to privacy. Although there are other default features such as~\textit{Siri} and~\textit{Google Assistant} which have privacy implications, we excluded them from the tasks list because they are not enabled by default. Nevertheless, we still asked our participants if they had these features enabled in their own devices. We chose the following tasks:

\textbf{\textit{Disable location services.}}\\
We asked participants to locate and disable the~\textit{location services} on the smartphone. Since location service is one of the most demanded services on smartphones~\cite{balebako2013little}, our aim was to investigate if users can locate and disable this feature. We then asked them the privacy implications of disabling this feature, and if they have it enabled on their phones. For iOS users, we further asked them if features such as~\textit{Find my iPhone} would continue to work properly.

\textbf{\textit{Restrict App from using a default feature.}}\\
To understand whether users understood the implications of MPDF settings on other applications (i.e., downloaded third-party and platform apps), we asked participants to disable a default feature on the device to restrict other apps from accessing it. In iOS, a~\textit{Camera} application defaults to tagging GPS information on all photos. Therefore, in this task we asked our iOS participants to restrict the~\textit{Camera} App from tagging their photos with location. However, since Android implements this feature differently, we asked our participants to restrict their~\textit{Google} App from having access to their~\textit{Calendar} and~\textit{Location} services.

\begin{figure}
	\centering
	\subfloat[]
	{ \includegraphics[width=0.45\columnwidth]{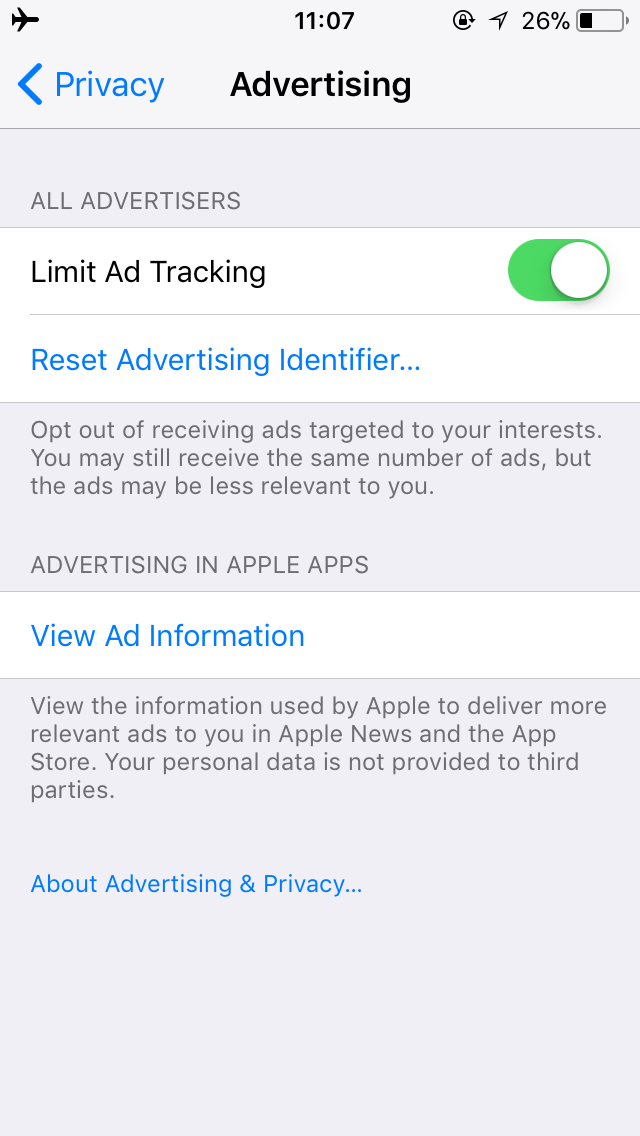}
		\label{iOSAdtracker}}
	\hspace{0.19cm}
	\subfloat[]
	{ \includegraphics[width=0.45\columnwidth]{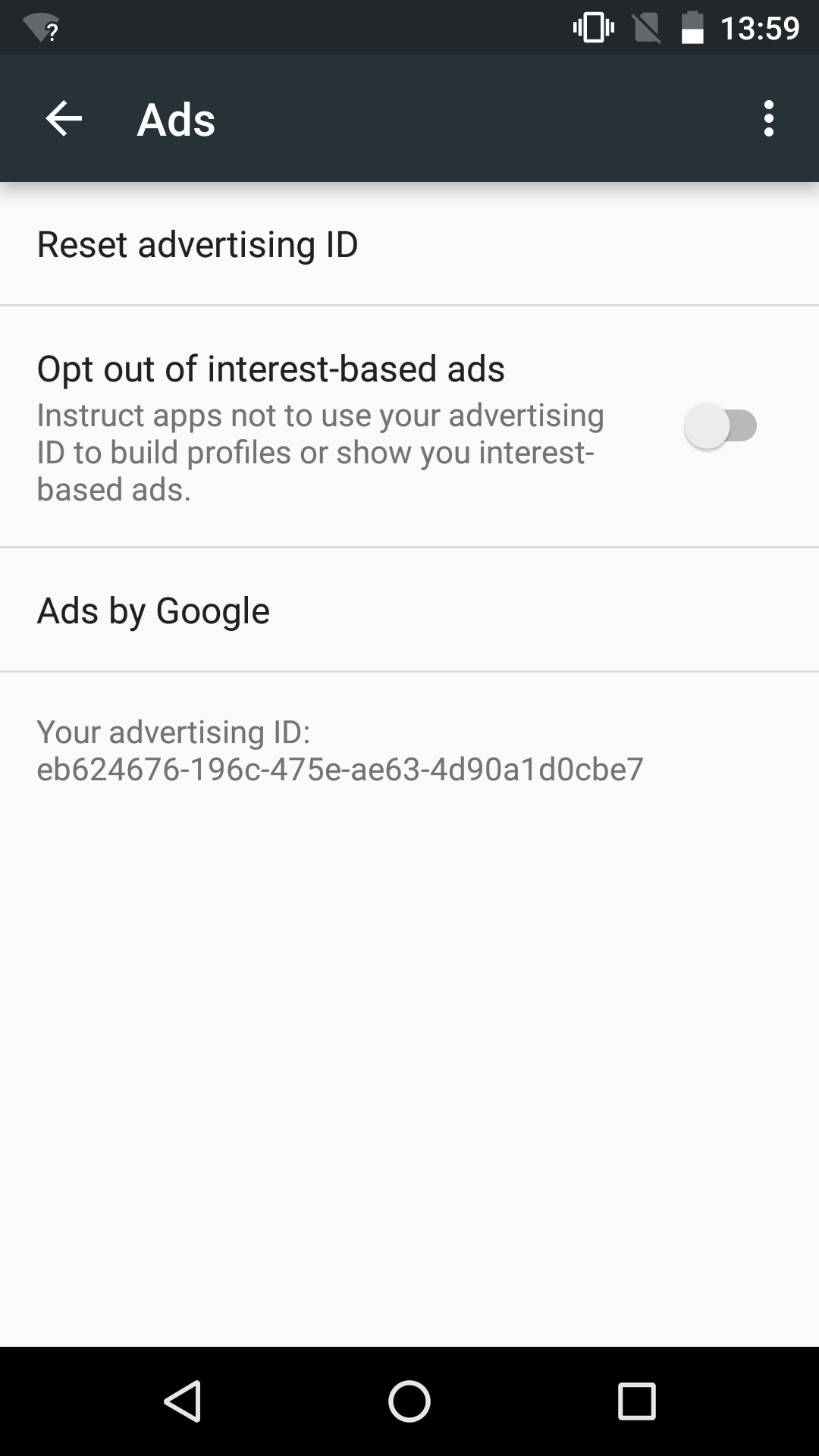}
		\label{AndroidAdtracker}}
	\caption{(a) iOS Ad-Tracker controls (b) Android Ad-Tracker controls}
	\label{fig:adtracker}
\end{figure}

\textbf{\textit{Disable/Limit ad tracking.}}\\
We asked participants whether they received adverts on their smartphones regularly and if they monitored or regulated the ads that they received. Our focus question then required participants to disable or limit~\textit{ad tracking} on the device provided. The purpose of this task was to investigate if participants could locate and adjust this feature and to find out if they understood the impact of MPDFs on other third-party applications. Figure~\ref{fig:adtracker} shows default settings for both platforms.

\begin{figure*}[!htp]
	\centering
	\includegraphics[scale=0.47]{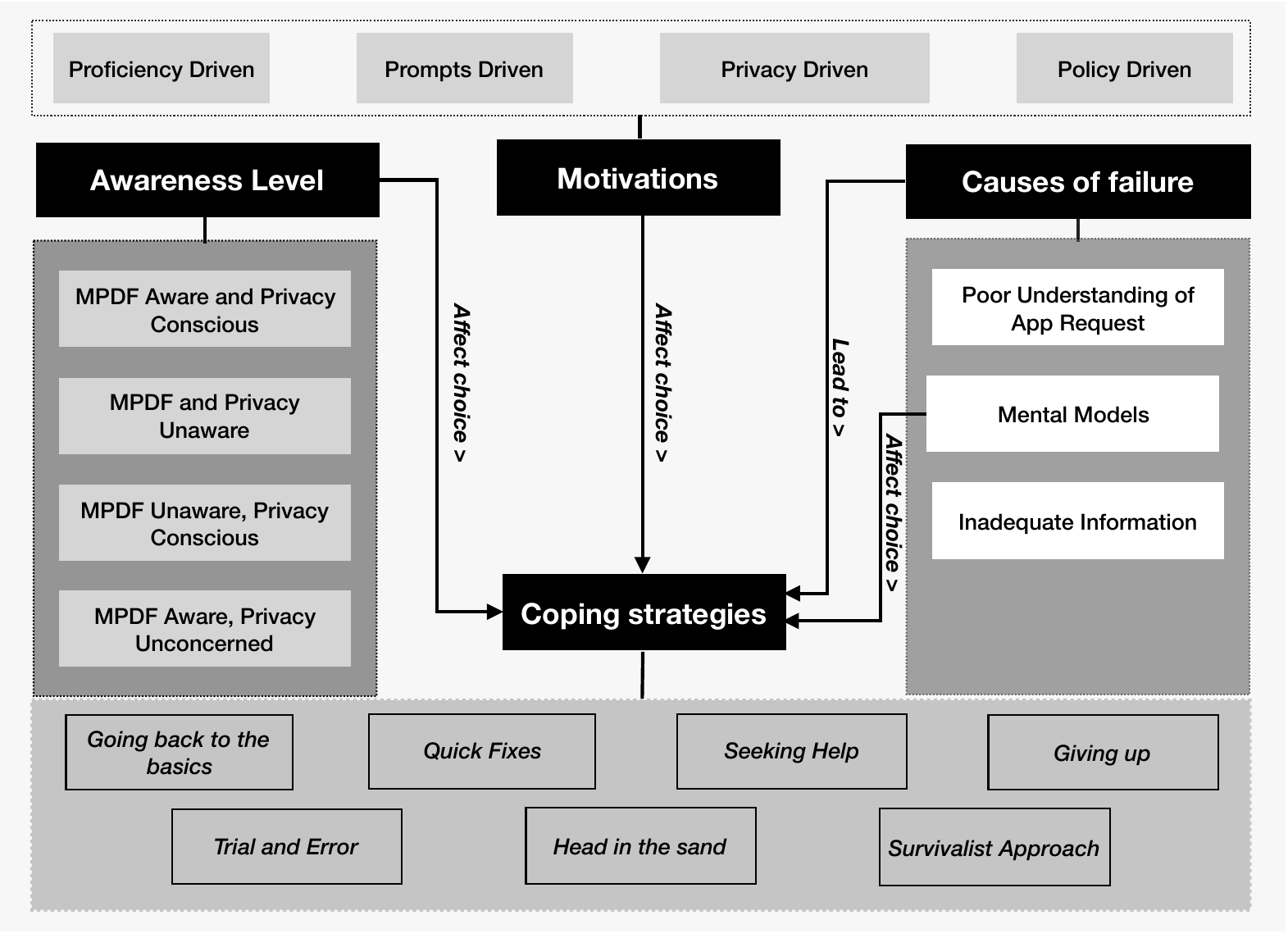}
	\caption{Key Findings: Illustration of how key findings (awareness levels, motivations and coping strategies) are related to each other. When users encounter some challenges while attempting to configure MPDFs, they adopt some coping strategies. Their choice of coping strategy is dependent on the type of setback, their perception of MPDFs, how data flow within their smartphone, and their motivation to configure MPDFs.}
	\label{Fig:summary}
\end{figure*}

\textbf{\textit{Restrict Usage and Diagnostic Report (Android only).}}\\
We asked our Android participants how often they used interactive options available to communicate with software app designers on their smartphone. We intended to investigate whether users were aware of and open to sharing app data with developers. Thus, we asked participants to locate and restrict~\textit{Diagnostic and Usage report} on the smartphone.

\textbf{\textit{Disable analytical data sharing (iOS only).}}\\
To understand iOS users' willingness to share their personal data with platform developers intentionally, we asked them to locate and disable the sharing of the device's analytical data. In the previous versions of iOS (e.g., 10.2), this feature was labeled~\textit{Diagnostics \& Usage}. Asking users to carry out this task also gave us an opportunity to understand whether users felt conflicted, and to what degree, when deciding to share data with the provider to improve the platform performance.
\vspace{-4mm}

\subsection{Thematic Analysis}
\vspace{-3mm}
Data analysis commenced after the first five sessions: audio files were transcribed, and notes were consolidated, then we started thematic coding. The first five transcripts included both Android and iOS user responses. After we familiarized ourselves with the data, two researchers independently generated a list of codes. The two researchers then met and consolidated the most noticeable codes and themes into a shared codebook~\cite{braun2006using,ryan2000data}. We aimed to find relationships between each group where possible, noting similarities that were observed and of an important distinction. The initial coding revealed several interesting codes which were consistent in each group of users of a brand of smartphone. However, as the two researchers continued with grouping, we started to notice patterns that cut across both platforms. Then, we began to merge and outline clear definitions from our coding scheme and allocate names to each distinct theme. One researcher then continued to code the rest of the scripts while the second researcher just coded another five to ensure reliability. Our Cohen's Kappa inter-annotator agreement was 0.89, which is acceptable for qualitative studies. Where there were disagreements about codes and themes, we redefined the code or the theme that was contested. In the end, we generated a ``thematic map'' of our data analysis (See Fig.~\ref{Fig:summary}).
\vspace{-4mm}

\section{Key Findings}
\label{Label:Findings}
\vspace{-3mm}

Our results are purely qualitative but to provide context we will sometimes report on the number of participants in a given category. Fig.~\ref{Fig:summary} presents an overview of our key findings. We next detail each of these findings:

\begin{table}[!ht]
	\small
	\centering
	\caption{Awareness Groups: Most participants had encountered some MPDFs before, but those who reported knowing about them were mostly aware of common ones such as location services and personal assistants (i.e., Siri and Google Assistant). Most of them (11) claimed having never thought that they are sharing their data with the platform provider.}
	\renewcommand{\arraystretch}{0.9}
	\begin{tabular}{p{5.7cm} >{\centering\arraybackslash}p{2.0cm} >{\centering\arraybackslash}p{0.6cm}}
		\textbf{Awareness Group} & \multicolumn{1}{c|}{Android} & iOS \\ 
		\hline
		MPDF Aware and Privacy Conscious users  & \multicolumn{1}{c|}{3} & 3 \\
		MPDF Aware with limited concern & \multicolumn{1}{c|}{7} & 4 \\
		The Unaware & \multicolumn{1}{c|}{3} & 3 \\
		MPDF Unaware but Privacy Conscious  & \multicolumn{1}{c|}{1} & 3 \\
		\hline
	\end{tabular}
	\label{tab:awarenessgroups}%
\end{table}
\vspace{-6mm}

\subsection{Awareness Level}
\vspace{-3mm}

Our analysis revealed four categories or levels of awareness of MPDFs among participants: the aware and privacy-conscious; aware with limited concern; the unaware; and MPDF unaware but privacy-conscious. These groups corroborate and enrich those found in prior work  (e.g., Lin et al.~\cite{lin2014modeling}).

Users learn about MPDFs from different sources. Out of the 27 participants we recruited, thirteen (13) mentioned that they learned about MPDFs from social media and blogs posts while five (5) reported that they were made aware of MPDFs by their family members, friends and work colleagues. We found that in most cases, people are only made aware of the existence of MPDFs ordinarily for functionality purposes (i.e., they get to know about some of these features when wishing to accomplish a task) but not for privacy implications of such controls.
\vspace{-5mm}

\subsubsection{MPDF Aware and Privacy Conscious users}
\vspace{-3mm}

This is a category of participants who showed knowledge about MPDFs and had a high level of privacy awareness. Most respondents from this category were conversant with smartphone MPDFs, and they were able to distinguish these features from other settings. They were able to locate and adjust MPDFs when asked. One participant said, 

``\emph{...I am completely paranoid with sharing my data ... when I first got my phone the first thing I did was disable most of my default features because I felt they were too intrusive... only turn them on occasion when I had the need...''} P4

Most of these users learn about MPDFs from online platforms such as social media and blogs posts, while one reported that they were made aware of MPDFs by a family member.
\vspace{-5mm}

\subsubsection{MPDF Aware with limited concern}
\vspace{-3mm}

Our study revealed a group of users who were aware but with limited concern for privacy. This group had some knowledge of MPDFs but did not give it enough consideration when it came to safeguarding their sensitive data. We further identified two subcategories of users within this group. The first included users who were aware but mostly passive when it came to privacy.  
They bought into the popular clich\'{e} that their data was going to be taken anyway whether or not they made any effort to protect it. For example, P5 and P21 said respectively: 

``\emph{...I have done some research as regards manufacturer features and privacy...but I can imagine that most people are like me...who try to make their lives easier by not caring too much about privacy.''} 

``\emph{...in my interaction with many smartphone users, privacy and security are the least of their concerns, so long as the phone can deliver the functions they required...''}

The second subcategory of users was aware but had not taken time to understand the privacy implications of MPDFs or how to carry out privacy-related configurations. P11 said: 

``\emph{I am from computer science...and still there are things I don't know...we are not really aware of the things we should think about...you might think of obvious things like the location... but what about ad tracker.  That didn't cross my mind...''}
\vspace{-5mm}

\subsubsection{The Unaware}
\vspace{-3mm}
Our analysis also revealed a group of users who were genuinely not well informed about MPDFs and their implications on user privacy. Furthermore, this group did not have any understanding of privacy controls. They ideally agreed that their motivation for acquiring and using smartphones stemmed from many reasons other than privacy such as vogue, aesthetic features of the smartphone and status (mainly iPhone users). Some of these participants admitted that the closest they came to be aware of privacy and its implications (or that they could be affected) originated only from recent privacy and security incidents not necessarily related to smartphones, like WannaCry\footnote{http://malware.wikia.com/wiki/WannaCry}. Although these users are always buying new smartphones, they do not consider MPDFs or their privacy implications. For instance, P12 and P15 said respectively:

``\emph{A lot of us just don't really know... all this is technical... I always run away from what I don't understand...''}

``\emph{If there is something new (smartphone) I want to buy it...I don't really care about privacy''} 
\vspace{-4mm}

\subsubsection {MPDF Unaware but Privacy Conscious}
\vspace{-3mm}

Some users knew very little about MPDFs and their privacy implications. However, their responses during the interview suggested that they were concerned about privacy. They did everything they could to protect themselves, for instance, they concerted their efforts to peruse some aspects of terms and conditions to be secure. However, because of their lack of awareness of MPDFs, they ended up sharing private data with platforms unknowingly. During the task-based exercises, they managed to complete some tasks successfully through trial and error. After completing the study tasks, they quickly altered default settings in their personal smartphones.

``\emph{I feel so annoyed that they [manufacturers] have data on me that I didn't consent to in the first place... It is not great that apps have information about you.''} P2
\vspace{-4mm}

\subsection{Motivations}
\vspace{-3mm}

Our analysis revealed that users' motivations to adjust MPDFs were based on their own proficiency, app request prompt, policy, and privacy concerns.
\vspace{-3mm}

\begin{table*}[!ht]
	\small
	\centering
	\caption{Mental model distribution among participants. Multiple mental models may be held by a participant at the same time.}
	\renewcommand{\arraystretch}{0.9}
	\begin{tabular}{p{7cm} >{\centering\arraybackslash}p{2.9cm} >{\centering\arraybackslash}p{2.4cm} >{\centering\arraybackslash}p{0.6cm} >{\centering\arraybackslash}p{0.6cm}}
		\hline
		\textbf{Mental Model} & \textbf{Affect user's } & \textbf{Used as a coping } & \multicolumn{2}{c}{No. of Occurrences}  \\
		& decision to configure & Strategy & \multicolumn{1}{c|}{Android} & iOS \\ 
		\hline
		- Functionality first, settings later & $\checkmark$ & & \multicolumn{1}{c|}{13} & 11 \\
		- The internet has all the answers & & $\checkmark$ & \multicolumn{1}{c|}{9} & 11\\
		- Privacy as a physical security attribute & $\checkmark$ & &\multicolumn{1}{c|}{9} & 8  \\
		- Mobile configurations are easy to manage & $\checkmark$ & & \multicolumn{1}{c|}{7} & 9 \\
		- Providers are trustworthy & $\checkmark$ & $\checkmark$ & \multicolumn{1}{c|}{4} & 11 \\
		- Laissez Faire attitude & $\checkmark$ & $\checkmark$ & \multicolumn{1}{c|}{8} & 5 \\
		- Zero check: Privacy not a concern at initial set up. & $\checkmark$ & $\checkmark$ & \multicolumn{1}{c|}{5} & 8  \\
		- Privacy is security & $\checkmark$ &  & \multicolumn{1}{c|}{9} & 3 \\
		- One-time configuration & $\checkmark$ &   & \multicolumn{1}{c|}{4} & 6 \\
		- No prompt, no need for configurations & $\checkmark$ & $\checkmark$ & \multicolumn{1}{c|}{6} & 3 \\ 
		- Configuration breaks functionality & $\checkmark$ & $\checkmark$ & \multicolumn{1}{c|}{6} & 2\\
		\hline
		
	\end{tabular}
	\label{tab:mentalmodels}%
\end{table*}
\vspace{-4mm}

\subsubsection{Motivated by Proficiency}
\vspace{-4mm}

\textbf{The experts.}
We found that some users adjust their MPDFs because they know and understand the implications of these features – where the settings are or could be located and how they work. They are experts, and conversant with the settings by intuition though they notified us that their own devices were different from the one used in the study. They configure MPDFs all the time, disabling and enabling as they need and wish. 

P21, an ``expert'' explained,
``\emph{Manufacturers want to get as much data from users in order to improve their services and for the allocation of relevant adverts to interested users to generate revenue...they will take every opportunity to get data in return...I use Siri but not frequently...not until recently did I begin to use iCloud...I do however use a UC browser which has an ad blocker...I find websites these days asking me to reload or disable my browser that does not support arbitrary adverts...}''

\textbf{The overconfident.}
Since we asked participants to rate their adeptness in configuring smartphones, we aimed to find out if users who highly rated themselves could complete the study successfully. During the exercises, we noticed that some users failed to locate and configure these settings though they had perceived themselves as experts. While some overestimated their self-efficacy and failed to complete some tasks, we found that sometimes being overconfident encourages people to explore their phone features more.

\textbf{The non-tech savvy}
Some users who rated and reported having low skill levels in configuring settings stated they do sometimes ``fiddle around'' with their smartphones to check what settings are available. However, they confessed that they never try to change anything. 
\vspace{-6mm}

\subsubsection{Motivated by App Request Prompt}
\vspace{-4mm}

A group of our participants informed us that sometimes they adjust their MPDFs because they get prompted by some apps to do so. Our analysis revealed that this category of users either ignored these nudges or simply granted all app permissions in order to continue using the service functionalities without much concern for what the app requested to access.

P2 said,
``\emph{...unless it pops up...then would I eventually realize that I need to look at data setting...}''
\vspace{-9mm}

\subsubsection{Motivated by Mandatory Policies}
\vspace{-3mm}

Some participants were driven to consider MPDFs only when they were constrained to do so by policies. The mandatory setup policies embedded in smartphones often compel users to make decisions on MPDFs -- they are forced to either leave them enabled or not use the feature on the phone at all. For instance, before Android version 6, Android offered users a choice to either accept all app request or delete the app. If users denied the request, then they would lose all the app functionality. Users reported having to use MPDFs because their platform did not give them a choice, e.g., being forced to use iCloud credentials to set up an iPhone.

``\emph{...how was I supposed to continue... I had to... I didn't want to break my phone.}'' P7
\vspace{-6mm}

\subsubsection{Motivated by Privacy Concerns}
\vspace{-3mm}

Some participants adjust MPDFs out of concern for their sensitive data. This group overlaps with those that adjust their MPDFs out of expertise, in the sense that they adjust their MPDFs because they understand the implications of MPDFs. Many people in this group simply turned off all possible features that made their personal data sharable or accessible to third-party apps or the platform. Our participants informed us that this came at the cost of functionality and usability.

``\emph{...I know about these features. I went through my phone and disabled everything. I care about my privacy.}'' P4
\vspace{-4mm}

\subsection{Causes of failures}
\vspace{-4mm}

Our results suggest that limited understanding of app request functions, insufficiency of information, and incomplete mental models impede users from adjusting their MPDFs. 
\vspace{-4mm}

\subsubsection{Limited understanding of App requests and permission implications}
\vspace{-3mm}

Some interviewees attributed their difficulty to correctly configure MPDFs' settings to misinterpreting app request. Our analysis also suggests that, during initial setup, users are bombarded with these permissions which they struggle to understand. Consequently, they end up being ambivalent about the implications of enabling or disabling MPDFs.

``\emph{...people just buy smartphones but think that they do not necessarily need to know about its [sic] functions...the features may be there, but they don't understand the settings...}'' P10
\vspace{-6mm}

\subsubsection{Insufficient information}
\vspace{-3mm}

There is insufficient MPDF information on smartphones to help users make informed decisions on whether they should have them enabled or disabled. Also, our participants revealed that there is far more information about the benefits of MPDFs than there is information about their implications from a privacy perspective. But, some users suggested that there is information available, nevertheless it does not relay the entire message clearly. For instance, one participant said:

``\emph{...not understanding the meaning of the exact terminology manufacturers use and not being aware of what some terminologies infer... for instance, <manufacturer> declares that they collect essential primary data to keep your software up to date and help improve Motorola products... but they are not saying it as it is, \textbf{what do they actually mean by help when they are stealing my data}...}'' P17
\vspace{-4mm}

\subsubsection{Mental Models}
\vspace{-4mm}

Prior research on mental models has suggested that users often fail to protect themselves or complete security and privacy tasks due to their limited technical understanding of systems~\cite{wash2010folk,wash2011influencing,ramokapane2017}. We, therefore, sought to understand smartphone users' general mental models with regards to MPDFs. Our mental model analysis was based on whether participants' mental models played a role in the following four aspects: (1) their decision to alter settings of a default feature, (2) how they carried out the configuration tasks, (3) their coping behaviors, and lastly, (4) we wanted to know if both iOS and Android users held a particular mental model. We derived these mental models from observing how users completed their tasks and what they reported during the interview. Table~\ref{tab:mentalmodels} provides a summary of these mental models.

\textbf{Functionality first, settings later.}
Our analysis revealed that most users prioritize functionality over privacy, especially when they are required to carry out some configuration task. We found that there is almost a trade-off between the two. However, most users will opt to use a feature and then change settings later. Some participants from this group considered configuring MPDFs as a waste of time and a hindrance. Most users who own this mental model often leave their MPDFs unchanged at the initial setup of their smartphones. We also found that this belief is shared among users who considered the process of configuring MPDFs and privacy controls either overwhelming or too cumbersome. 

``\emph{Sometimes you have to compromise certain aspects of your privacy for an app function like me...}'' P11

\textbf{The internet has all the answers.} 
To cope with some default feature configuration challenges, most users adopt a ``Internet has all the answers'' model. This is a belief that search engines know what they need to do. When users encountered a challenge, they would search for their solution online. We found that users would rather choose to search through Google than to read their service agreements or policies. This mental model co-exists between all smartphone users and contributes to users changing their default feature settings. Some users use this mental model as a coping strategy. 

``\emph{No one reads a phone manual...people tend to search Google and other search engines rather than actually checking the phone manufacturers information...}'' P11

\textbf{Privacy as a physical security attribute over privacy inclusive of back-end protection. }
Many of our participants understand privacy as a physical security attribute and believe that it had nothing to do with making sure that the platform does not share data unauthorized. Many users concert more effort securing their data from physical access rather than through misconfigured features. When asked what they do to preserve their privacy, they stressed using long passwords, screen locks, and biometric locks. This was a shared belief between the iOS and Android users and mostly within the Unaware group. They exhibited limited awareness of the data sharing ecosystem that existed in their smartphones.

``\emph{...I look at privacy from the angle of... can one get into my phone rather than can manufacturer steal my data.}'' P7

\textbf{Mobile configurations are easy to manage. }
We observed that after completing the first task (disabling location settings), many participants trivialized the rest of the tasks. We had intentionally started by asking participants to disable location. We aimed to give users a task they may have completed before, therefore helping them to relax while taking the study. Balebako et al. suggested that most users know how to configure location settings~\cite{balebako2013little}, this was confirmed in our study. However, we noticed a change in confidence as the task changed and required them to configure features of which they had hardly heard. While some users initially struggled but then managed to complete some tasks, they admitted that they had underestimated the tasks. When asking one participant why they failed to disable \textit{ad tracking}, they responded:  

``\emph{I couldn't find it... I think it is a lot of commitment to configure settings.}'' P5

``\emph{I will try, but now I don't think I actually know how... I just never thought about this.}'' P11

\textbf{Providers are trustworthy. }
Some users do not alter the default features because they trust in the platform provider's data practices. They believe that the status of their chosen platform speaks for itself and therefore they do not need to worry about their data and privacy. They believed their platform was secure and impenetrable hence could not share data without their knowledge. We found this belief to be common among iOS users, especially those who were aware of MPDFs but not concerned about privacy, and those who were unaware of MPDFs. When asked whether their default features could share their sensitive data unknowingly, P13 said:

``\emph{No, I trust the device and its applications...}'' 

\textbf{Laissez Faire Attitude. }
When some users face difficulties in adjusting or understanding the privacy implications of some MPDFs, they resort to giving up with the belief that the platform provider can do whatever it wants. Users with this mental model explained that they did not mind if platform providers harvested and processed their data without their knowledge. However, users who were aware of default features had high privacy concerns and had the expertise to adjust default settings, were less likely to adopt this strategy. We found this strategy to be popular among both platform users. 

``\emph{...I take a laissez-faire approach, constantly believing everyone is being spied on, so I don't care as much... I don't think I'm that important...}'' P27

\textbf{Zero check: Privacy is not a concern at initial setup. }
Some users do not consider privacy at initial setup. They believe that privacy is an issue that should be revisited later on when they have used the phone. When asked whether or not they gave reasonable consideration to the MPDFs with regards to privacy at initial setup, one participant said,

``\emph{...you buy a fancy phone with fancy graphics and you think that's fine, I don't need to check anything...}'' P14

This was common among users of both platforms. 

\textbf{Privacy is security. }
Some of our participants faced challenges differentiating privacy from security~\cite{Hadar2018}. While some confessed they did not understand the difference, we observed during the study tasks that when asked to configure a privacy setting, they would usually search under the security menu. We identified this with most of Android users. We infer that this may be because Android has a security label while iOS does not. However, some iOS users do have this model as well. While attempting to disable location services, P8 said:

``\emph{...now I'm going to settings and security...let me check... it's going to take me a while to find it...}''

\textbf{One-time configuration. }
Some users consider privacy related configurations as a one-time process. They believe that having disabled a feature at an initial setup that feature will remain disabled until they actively enable it again. Most of these users never revisit their default setting screens to check if they still have their desired settings. However, OS upgrades and some applications may alter their initial settings, for example, the recent iOS upgrade defaulting back to enabling iCloud storage for users who had disabled it before the upgrade.

\textbf{No prompt, no need for configurations. }
Users choose to react to prompts rather than to pre-emptively configure their settings. When initially setting up their smartphones, some users -- the unaware group -- normally bypass the setup process of these features. Or they configure them without much consideration of the impact of their decisions. They do not revisit their settings to confirm if they have been correctly configured -- with the mindset that if there is any need to change them, they will get a prompt requesting them to do so. Some participants even revealed that they could not consider the settings of a feature that they do not know or that has not been run through during initial setup. Hence, they will set it up when the need arises which is usually through prompts. We found this mental model to exist in both iOS and Android users. Some Android users said,

``\emph{I didn't even know there was an ad tracker.}'' P1

``\emph{...Unless it pops up would I then come to the realization that I need to look at data settings...}'' P2

When asked whether their iPhone could send or download data without their knowledge some responded:

``\emph{I don't know if it can without my approval... can it?}'' P6 

``\emph{when my phone tries to download things, I always receive a notification and refuse...}'' P23

\textbf{Configurations break functionality. }
Some users own a belief that altering default settings and features will break their phone's functionality. This results in them not changing their default settings. We found this to be common among participants who were less aware and informed about privacy. For some users, it was due to negative past experiences. We observed that users occasionally adopt this ``thinking'' as a coping strategy especially when they do not understand the implications of disabling a certain feature. This is shared by both iOS and Android users. One user said:

``\emph{I hope that I do not end up breaking the phone...that is why I do not alter my device...}'' P7

``\emph{...when you disable the manufacturer apps you won't be able to get (retrieve) the pictures... it's a bit weird...}'' P14
\vspace{-6mm}

\subsection{Coping Strategies}
\vspace{-4mm}
Users tended to create different coping strategies to help palliate their inability to complete the configuration of MPDFs and their privacy-related controls correctly. For easy explanation, we have categorized and named these strategies. When users fail to locate the settings of default features, it is common for them to pause and revisit their basic approaches to adjusting settings. It is also common for users to employ quick fixes and avoid asking for help when they think it is something they should know. Three (3) of our participants gave up while attempting to locate MPDF controls. However, we found that privacy-conscious users rely on other people to configure MPDFs. In the following, we discuss these strategies in detail.
\vspace{-8mm}

\subsubsection {Going back to the basics}
\vspace{-4mm}

While most of our participants had high self-rated expertise, during the one-to-one interviews, we observed that some might have over-rated their expertise. When they encountered some challenges locating certain settings, some would pause and reconsider their approaches. Most would eventually complete the task correctly but with much cautiousness after first failing to do so. On further examination, we recognized three possible reasons why users adopt this:

\begin{itemize}
	\item  Users either underestimated the difficulty of the tasks or assumed they understood features more than they actually did. Users who were aware of MPDFs but with limited concern mostly encountered this problem.
	\item  Users placed illogical reliance on dogmatic approaches they may have used before to accomplish similar privacy configuration tasks. For instance, they may have trusted some app permissions which were convenient without considering their privacy implications, e.g., Siri.
	\item Failure to re-evaluate. Users did not think it was necessary to revisit their settings after downloading third-party apps that required the use of location. These kinds of apps often lead to location service continually being enabled, leaving users just as vulnerable as those who fail to make the correct app permission decisions.
\end{itemize}
\vspace{-8mm}

\subsubsection{Quick fix}
\vspace{-3mm}

This category of participants was more comfortable with running searches (using search engines) to locate relevant MPDF controls. Some participants in this group also attempted to install some tools such as Mobile Device Manager Plus, especially those who were unsure of how to go about locating certain features. We observed that this group of participants were not necessarily the most aware regarding issues relating to privacy, nor were they the most conversant with MPDFs and privacy controls. However, because of their self-created shortcuts, they were able to configure and learn about MPDFs.
\vspace{-6mm}

\subsubsection {Survivalist approach}
\vspace{-4mm}
Some participants adopted a relatively withdrawn yet conscious approach we called ``Survival approach''. They still made some effort to complete the task but appeared to be a bit distant, either through overthinking a previous incorrect result or not giving much thought to their actions while carrying out the tasks. In this case, they mentioned guilt and fear from realizing how vulnerable they were for not being able to locate some of the MPDFs. We identify their behavior as (what we understand and assume to be) self-resignation to decisions they had incautiously made regarding certain app requests. They showed discomfort and regret when realizing that they were not familiar with or did not fully understand the impact of their actions on their privacy. To come to terms with the implications of these features and app requests that they had often trivialized in the past, they unconsciously adopted a survivalist approach of attempting to correctly complete the task but not necessarily thinking about what they were doing in the process. A handful of these participants did not complete the tasks successfully. We identified this among users who were privacy aware but lacked adequate awareness of MPDFs.
\vspace{-6mm}

\subsubsection{Head in the sand}
\vspace{-4mm}

This was observed in participants who were not aware of MPDFs and their privacy implications. When they failed to complete the exercise, they claimed that they find no reason to change their settings. They reported that they had nothing to hide and were less bothered about privacy infringements that MPDFs could cause. For instance, some users did not see any need to turn off location settings, even when they were not necessarily using any location-based application service. We identified this behavior across both platforms.
\vspace{-6mm}

\subsubsection{Seeking help}
\vspace{-4mm}

A number of our interviewees reported that they have asked for help disabling or understanding the implications of some MPDFs. They sought advice on what actions to take whenever they noticed strange activities on their phone. Seeking help was not only restricted to asking people, but it included searching online (see, e.g., Quick fix above). This is common among those who had a fear of being ridiculed for being ignorant. As most users learn about MPDFs from other people, they also seek help from those who make them aware. The most proficient users among our participants revealed that this approach had helped to increase their understanding and skills.

``\emph{...If you have friends or families that are aware or know about privacy and security...ask someone who has skill...}'' P24

Our analysis suggests that this \textit{socialization of the privacy configuration experience} gives users a greater chance of correctly configuring and understanding MPDFs. 
\vspace{-6mm}

\subsubsection{Trial and Error}
\vspace{-3mm}

Some participants became visibly frustrated when they could not perform certain tasks that would ordinarily require them to do an in-depth search under the privacy settings of the smartphone. They became overwhelmed at times and extended their search to the security options. One participant even considered the device storage application. When queried for erroneously going that far, the participant explained that the device storage icon looked attractive and aroused their curiosity to look for the privacy feature they assumed could be hidden under the device storage menu. Users who adopted this approach came from the category of participants who were unaware but privacy-conscious. P5 said,

``\emph{...it is a lot of commitment on users for manufacturers to expect us to know how to configure all privacy setting... there should be a video that can't be skipped for users to abide by to help them configure their privacy settings correctly. However, for those who do not watch it they'll end up like me...} 
\vspace{-6mm}

\subsubsection{Giving up}
\vspace{-3mm}

A few of our participants were identified to have a very low endurance threshold when faced with the pressure of not knowing how to locate and adjust certain features that could potentially affect their privacy. This group simply gave up after the first few attempts, accepting that they knew very little about MPDFs and never really considered privacy in this context. One participant from this group claimed to be less paranoid by not knowing about MPDFs and believed that what they did not know (i.e., privacy) would not hurt them. While some of these users were not concerned about privacy, some showed interest in knowing how to configure their MPDFs. When asked about their own MPDF configuration practices especially about how they deal with prompts, they reported that, in most cases, they ignored them unless ignoring them limited functionality. We found that participants from this category are likely to grant most of their app permissions in exchange for functionality.
\vspace{-6mm}

\subsection{Users' desires regarding MPDFs}
\vspace{-3mm}
Apart from a consensus on the need for users to do more towards gaining knowledge about MPDFs and their privacy implications, our participants raised some issues they considered pertinent for designers to consider with regards to MPDFs, user privacy, and smartphone interfaces. The four most common ones were: (1) reliable and clear policy, (2) clear outline of the implications of MPDFs, (3) MPDF configuration information, and (4) standardized menu interfaces.
\vspace{-7mm}

\subsubsection{Reliable and clear user policies}
\vspace{-3mm}

Like other previous studies~\cite{mcdonald2008cost}, participants shared their concerns about the uninspiring length of privacy terms and conditions although they are meant to spell out the salient points regarding smartphone functionality and app implications. Participants who were aware of MPDFs, privacy controls, and the impact of app permissions, were not so interested in smartphone policy guides but rather on the need for greater privacy controls allocated to users to safeguard their data from manufacturers and third-party apps. This idea was popular among users who were aware and conscious of privacy. 

``\emph{...I suggest a summary sheet of ten key privacy things sold with the device by designer...}'' P7
\vspace{-5mm}

\subsubsection{Unambiguous outline of the implications of MPDFs and relevant information}
\vspace{-3mm}

Many of our participants felt that there is a dearth of adequate information on MPDFs in general, and there is also not enough clarity on their implications. They suggested that manufacturers should adopt methods that would be more appealing, empowering, and proactive to inspire users to peruse policy information regarding MPDFs and their privacy implications. One of the participants recommended the use of short video clips that touch on matters relating to privacy adequately since people would be more attracted by visual than text-based content.

``\emph{...I suggest the best way is through adverts when they are marketing the phone...I think that's the best way to reach people...If the companies could also use a feature on their phone that pops up daily notifications that give information about it (privacy)...I don't think people are that ignorant not to want to try to understand what it's about...}'' P25
\vspace{-5mm}

\subsubsection{More ethical sincerity from app designers and service providers regarding user privacy}
\vspace{-3mm}

Some users expressed their concern towards platform designers whom they felt were not entirely sincere about user privacy. Some suggested that it would be of benefit to know how the collated data is being stored, used, shared, and deleted as these factors play an essential role while configuring their privacy settings.

``\emph{...there is a need for understanding the cost and benefit of allowing certain function...it's hard to tell if you turn off a certain feature what that will limit you access to versus what you are limiting others access to on your phone...}'' P27
\vspace{-5mm}

\subsubsection{Standardization of smartphone privacy interfaces}
\vspace{-3mm}

During our interviews, some users reported that it was difficult to find their way to the right MPDF settings due to the differences in smartphone user interfaces. They suggested that standardized and less complex interfaces could help them locate and configure MPDFs easily. P20 said

``\emph{I don't think it's clearly laid out, the privacy settings are all over the place...they could actually do more to categorize them... because of how they are spread about, it tends to undermine the importance...ordinary users would not take it as seriously when it's all over the place...they should do more...}''
\vspace{-6mm}

\section{Discussion}
\label{Label:Discussion}
\vspace{-3mm}

In contrast to previous research~\cite{iachello2005developing,chin2012measuring,felt2012android,felt2012ve,liu2014reconciling,Tsai205132,ismail2015crowdsourced,ismail2017permit}, our work revealed the following:

\begin{itemize}
	\item Usage or awareness of MPDFs does not necessarily imply that users understand the implications of using such features. Besides, we found that users cannot easily locate and alter these settings. 
	
	\item Users consider MPDFs (i.e., apps they haven't downloaded) as an integral part of the platform. As a form of trust, they were more cautious with downloaded apps than the ones they find in their phones by default.
	
	\item While users first encounter MPDFs during their initial smartphone setup, they are often overlooked and left unchanged. Most users will later learn about these features and settings from other people including family members, friends and other online sources. Our results suggest that, compared to other MPDFs, most people are aware of location services because of the requests they receive from apps that need to use location. 
	
	\item Users do not understand the data sharing ecosystem of MPDFs, that is, how these features work or what information they collect. For instance, some users do not understand that by using Siri their information will be shared with Apple and this may contain personal information. 
	
	\item Users attribute their challenges of configuring MPDFs to complex app requests, hidden MPDFs controls (e.g., Ads identifier) and lack of relevant information on their implications. Some features need the user to visit several configurations to disable them, for instance, to disable location tracking on Android, a user is required to visit several places (e.g., location control under settings and then activity tracking) to turn it off.
	
	\item Users' coping strategies vary greatly and depend on the challenge, how they were made aware of MPDFs, proficiency level, motivation, the time and situation. Nevertheless, both platform users deploy quick and temporary solutions. Unlike the study in~\cite{ramokapane2017}, users are not intimidated to ask for help. They reported that asking for help gives them the confidence they have in their desired settings, and they have the opportunity to learn more about these features including the ones not investigated in our study.
\end{itemize}
\vspace{-9mm}

\subsection{Android vs iOS}
\vspace{-3mm}
The majority of our participants who used the iOS platform fell into the category of those who were \emph{MPDF aware but unconcerned}. Our interviews revealed that most of the iOS users were only aware of MPDFs that they mostly use (e.g., Siri) but they were generally \emph{MPDF unaware} and did not know what data is shared with the platform provider through MPDFs. Nevertheless, most iOS participants were less concerned about MPDF privacy implications unlike some of their Android counterparts who were mostly \emph{MPDF unaware} but had the notion that their platform provider might be collecting information about them (\emph{MPDF Unaware but privacy-conscious}). Both platform users were not aware of Ads identifier, usage and diagnostic reports, and analytics. However, we found that both platform users were aware of location services because it is a feature frequently requested by other apps.

Most Android users were likely to turn off their settings than their iOS counterparts. When being made aware of MPDFs and their privacy implications, Android users were likely to disable. Some Android users reported having disabled them because they were not interested in Google knowing everything about them. Some iOS users argued that their analytics data is helping Apple improve their phones. However, when asked about \emph{Ads identifier} both platform users were interested in disabling it. 

During the study, Android users highlighted the differences in the interface or the settings of the phone we used for the study than the iOS users. They were more frustrated when struggling to locate specific settings than those who used the iOS. They ended up searching randomly for settings. Android users who were more knowledgeable about smartphone settings were able to navigate to find the settings although they used different phones than in our study. iOS users did not complain about the interface but rather, acknowledged they did not know where such features were located.

Android users were more reactionary than pre-emptive like iOS users. We assume this is because Apple had long implemented dynamic permissions. iOS users were more likely to trust their platform than Android users. iOS users believed that the platform was designed with their privacy in mind. This may be a marketing effect than a technical one.
\vspace{-6mm}

\subsection{The impact of Opt-out mechanisms}
\vspace{-1mm}
Opt-out mechanisms are created to exploit people’s behavioral and psychological practices to encourage data disclosure. Recently, for instance, Google has been found to track users' location even when location tracking has been turned off~\cite{AP_location_tracking}. By default, users’ location can be tracked through web applications or activities that use locations such as maps. In order for users to turn off location tracking, they are required first to locate these settings, however, they may not be aware that they exist in the first place. This highlights that manufacturers in some cases use these mechanisms to suppress privacy concerns~\cite{acquisti2015privacy}.

Our results affirm that people do not change their MPDFs for convenience reasons and, in most cases, they understand them as manufacturer’s recommendations. This conclusion may explain people’s privacy behaviors.

Some default features are designed or accompanied by explanations that confuse people and may lead users to prefer leaving them unchanged~\cite{conti2010malicious}. Some descriptions may make people think that the safest option is to leave them untouched hence leading to greater sharing (e.g., Google assistant opt-out controls). Privacy behavior studies describe this as uncertainties that people have about privacy~\cite{acquisti2015privacy}. Some  ``MPDF unaware'' users reported that, in most cases, there are never sure about the consequences of changing default settings and if this would affect how their devices work.
\vspace{-6mm}

\subsection{The role of regulation}
\vspace{-4mm}
Our results suggest that consent and choice do not necessarily provide enough control over people’s data. New regulations should aim to balance manufacturers’ interest in users’ data against what users may be willing to share. While platform manufactures may fear for fewer data disclosures, policies should encourage mechanisms that allow users not to share information by default. For instance, a policy that is similar to GDPR article~\footnote{http://www.privacy-regulation.eu/en/article-25-data-protection-by-design-and-by-default-GDPR.htm} 25(2) requires that the organization which collects data for processing should make sure that, by default, personal data are not accessible without individuals' intervention (i.e., a user should act to allow data collection). Article 25 also refers to the ‘by default’ theme by specifying that providers should only collect data that is necessary, thus enforcing data minimalization. Our findings, therefore, suggest that smartphone platform designers should refrain from using opt-out as default settings when setting up smartphones because not all the data collected by default may be deemed necessary for the functionality of the phone. It should be noted that there is no evidence that all the data collected through MPDFs are necessary for their functionality.
\vspace{-6mm}

\subsection{Early does not equal transparency}
\vspace{-4mm}
Displaying such controls to users early during setup may seem too desirable – as it may be interpreted as transparency. However, sometimes transparency mechanisms displayed at the time when users are still excited about their new devices may be ineffective. Presenting these controls at setup presents biases that even the most privacy-conscious may leave them unchanged, not considering the future consequences because immediate gratification of using the device always trumps the delayed one. Research has previously shown that showing users privacy policies (i.e., firm’s data practices) before they sign up for services is not beneficial because people do not read them. Social networking surveys~\cite{strater2008strategies,hargittai2010facebook, Misra:2017} have shown that privacy needs change over time (i.e., at one point in our lives we are privacy pragmatists, privacy fundamentalist, or privacy unconcerned~\cite{Westin:2001}). Hence privacy restrictions and disclosure decisions made earlier may need to change. 

Overall, the results we present show that only a handful of users can locate and configure MPDFs. However, for those who adjust them because of prompts, we suspect users get prompted merely because they had initially turned them off.
\vspace{-4mm}

\subsection{Limitations}
\vspace{-3mm}
Our study has several limitations. 
First, we adopted convenience sampling due to limited resources and geographical proximity, and we acknowledge that this method of sampling may introduce some hidden biases~\cite{etikan2016comparison}. In order to minimize biases, during the screening phase, after filtering out those who did not qualify, we divided the whole group into two broad groups, iOS and Android, then randomly invited people for interviews. However, it is possible that those who took part in the study were the ones who responded promptly, hence results were more inclined to their usage and perceptions. 

Second, during the study we were using a Motorola X ($2^{\text{nd}}$ Generation) running Android 6.0 and iPhone 6 running iOS 10.3. It is likely that unfamiliarity with the phone brands and OS versions we provided for the study could have influenced performance or caused confusion during the task-based exercises. Our results are also limited to these versions. However, people change phones and upgrade operating systems quite often. So lack of full familiarity with the interface is a realistic scenario encountered by users when adjusting MPDFs. For instance, our Android users may be using a phone from a different manufacturer which may have an interface that is slightly different from the one offered by Motorola. Also, our iPhone participants may have been using older versions of iOS than the one used in the study.

While our findings are helpful in recognizing the challenges and the perception of MPDFs, there are limitations to what questions our data can help us answer. Future work and privacy discussions should consider these limitations, e.g., while every smartphone user has one way or another set up their phone, we still do not know the extent to which they understood the configurations they made when setting up their smartphone or if their configurations at all reflected their intentions. Moreover, it is still not clear whether those who failed to locate the settings of these features were not aware of them, or even if they have not disabled them on their phones. It is possible that some users may have chosen their desired settings during setup, i.e., for MPDFs which are displayed to users during initial setup. Prior studies~\cite{strater2008strategies,hargittai2010facebook, Misra:2017} in social networks have found that sometimes users’ settings do not match their intentions or show some inconsistencies.
\vspace{-6mm}

\section{Conclusions}
\label{Label:Conclusion}
\vspace{-3mm}
We have studied the usability of configuring privacy of MPDFs in smartphones. Our study highlights the complexity faced by users in understanding the multi-faceted nature of the data eco-system in smartphone settings and the privacy implications of this eco-system. While there is improved awareness of data requested by third-party apps -- with the usage of dynamic permission models within iOS and Android -- the same is not true of MPDFs. With the exception of location services, our participants found it challenging to adjust the privacy settings for the MPDFs in the study. Issues ranged from inability to locate the MPDF in the first instance through to displaced mental models that focused only on password and PIN-based protection of phones and not the leakage of data via the various features inherent to the phone. The configuration of MPDFs during setup was also noted by participants as a secondary task with users opting to \textit{skip} or \textit{accept} the options on offer in order to start utilizing the phone; often not returning to adjusting those settings. This naturally begs the question: with the increasing incorporation of more and more MPDFs that gather user data in various forms -- from virtual assistants to integrated cloud storage and services -- are we approaching a new frontier for privacy in smartphone settings? Our study highlights the need for increased focus on usable privacy for MPDFs and mechanisms to \textit{genuinely} empower users in managing MPDF privacy. This is a non-trivial challenge that requires further studies and improved design of MPDFs.
\vspace{-5mm}

\section{Acknowledgements}
\vspace{-4mm}
We thank the anonymous reviewers for their suggestions and feedback. This work is partly supported by EPSRC Grants EP/P031838/1 and EP/N023234/1.

\bibliographystyle{abbrv}
\bibliography{main}

\section{Appendix}

\subsection{Interview Guide}
\label{app:Interviewguide}
Thank you for participating in our study. As you read in the consent form, we will be recording the session so we can review it to make sure that we don't miss any part of our conversation. Your information will be kept confidential and will only be accessed by us. Your name will not be associated with any data I collect. Do you have any questions regarding the consent form? Do I have your permission to start the recording?
\begin{enumerate}
	\item What kind of smartphone do you use?
	\begin{itemize}
		\item Follow-up-1: How long have you been using it?
		\item Follow-up-2: Did you set it up yourself?
		\item Follow-up-3: Would you say it is easy?
	\end{itemize}
	\item Do you consider yourself to be privacy conscious?
	\item How do you keep yourself safe from sharing your personal information through your smartphone?
	\begin{itemize}
		\item Follow-up-1: What kind of information are you currently not sharing?
	\end{itemize}
	\item Do you use the following apps? \\
	(If iOS)
	\begin{itemize}
		\item Siri
		\item iCloud
		\item iTunes
		\item Maps
		\item Camera app
	\end{itemize}
	(If Android)
	\begin{itemize}
		\item Google Assistant
		\item Maps
		\item Camera app
	\end{itemize}
	\item Have you ever restricted apps from accessing your location?
	\begin{itemize}
		\item Follow-up-1: Why?
		\item Follow-up-2: Did you do this by yourself?
		\item Follow-up-3: Can you show me how you would restrict an app from accessing your location?\\ 
	\end{itemize}[Encourage the user to share how they do it]\hfill \break
	[Note: If the user experiences some challenges, ask the user to share them.]
	\item Can you give me an example of a default feature or app you know?
	\begin{itemize}
		\item Follow-up-1: Do you know how to configure this feature?
		\item Follow-up-2: [If iOS] Can you disable your camera app from tagging photos with location. 
		\item Follow-up-2: [If Android] Can you please restrict your Google app from accessing your calendar information.\\
	\end{itemize}     
	[Encourage the user to share how they would do it] \hfill \break
	[Note: If the user experiences some challenges, ask the user to share them.]
	\item	Have you ever heard of ~\textit{ad tracking}?
	\begin{itemize}
		\item Follow-up-1: Where? What is it?
		\item Follow-up-2: Do you have it enabled in your phone?
		\item Follow-up-3: How does it work?
		\item Follow-up-4: Can you please disable ~\textit{ad tracking} from the mobile phone provided? \hfill \break
	\end{itemize}
	[Encourage the user to share how they would do it]\hfill \break 
	[Note: If the user experiences some challenges, ask the user to share them.]
	\item	Have you ever experienced problems with your phone? For example, your phone freezing during use.
	\begin{itemize}
		\item Follow-up-1: What was the problem?
		\item Follow-up-2: Have you ever shared crush reports with your phone provider?
		\item Follow-up-3: [If Android] Do you know that by default your phone shares this information with providers?
		\item Follow-up-3: [If iOS] Do you know about diagnostic and usage reports?
		\item Follow-up-4: [If iOS] Can you please disable analytical data sharing on the phone provided?         
	\end{itemize}
	[NOTE: If the user experiences some challenges, ask the user to share them.]\hfill \break
	[Explain to the interviewee that the task study has ended.]
	\item How do you feel about the tasks, you have just completed?
	\item What do you think should be done about this?
	\item What do you think providers should do to help people understand MPDFs? \hfill \break
	[Explain to the user that you are at the end of the interview, ask them if they do have any questions or anything they want to share about MPDFs.]
\end{enumerate}
\begin{figure}
	\centering
	\subfloat[]
	{ \includegraphics[width=0.45\columnwidth]{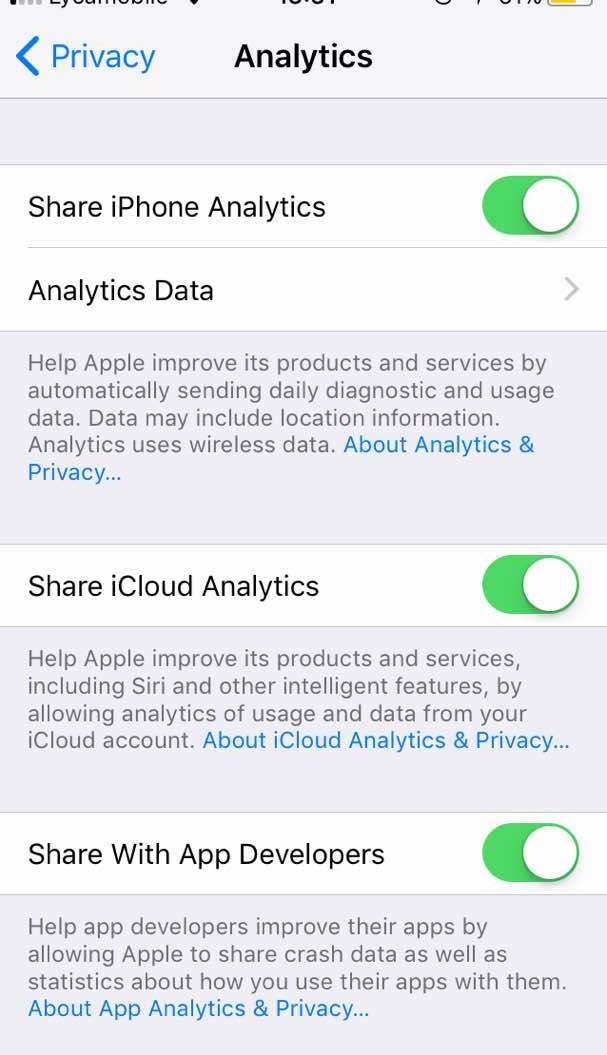}
		\label{Analytics controls}}
	\hspace{0.17cm}
	\subfloat[]
	{ \includegraphics[width=0.45\columnwidth]{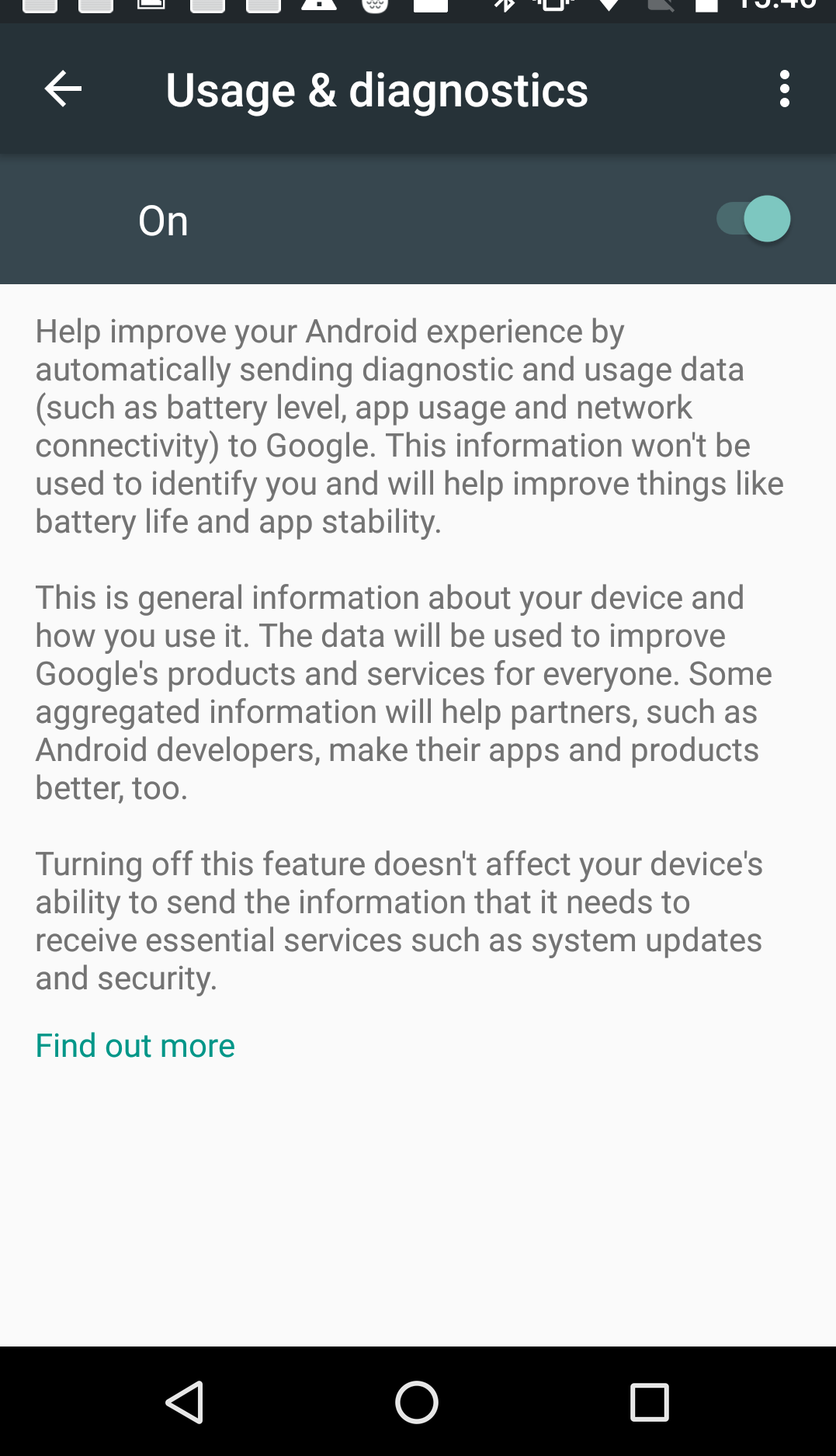}
		\label{Diagnostics and Usage}}
	\caption{Sample screenshots (a) iOS Analytics controls (b) Android Usage \& Diagnostics controls}
	\label{fig:diagnostics}
\end{figure}

\end{document}